\documentstyle[12pt]{article}
\catcode`\@=11 \@addtoreset{equation}{section}  
   
\title{Why we can not see the curvature of the quantum 
state space?}
\author{P. Leifer}
\date{Mortimer and Raymond Sackler Institute of Advanced Studies \\ Tel-Aviv University, Tel-Aviv 69978, Israel \\
e-mail:leifer@ccsg.tau.ac.il}
\begin{document}
\maketitle
\begin{abstract}
A new type of gauge quantum theory (superrelativity) has been proposed. This differs from ordinary gauge theories 
in sense that the affine connection of our theory is constructed from first derivatives of the Fubini-Study metric tensor. That is we have not merely analogy with general relativity but this construction should presumably provide a unification of general relativity and quantum theory. 

Here we shall discuss the physical meaning of metric
properties of the projective Hilbert space and 
manifestation of its nontrivial physical character.

\end{abstract}
\section{Introduction}
Gauge fields of modern gauge theories can not be 
expressed in terms of more fundamental fields [1]. 
In my previous works [2-4] we dealt with a geometrical
approach to the unification of quantum theory and
general relativity.  This based on the tangent fiber
bundle over complex projective Hilbert space $CP(N)$
of the pure quantum states. The linear connection (Christoffel symbol) in this theory as well as
connection in general relativity expressed in terms of
field of metric tensor. But now one should mean a metric
relationships not in spacetime but rather in space
of the pure quantum states, in the projective Hilbert
space equipped with the Fubini-Study metric.

If one wishes to endow the projective Hilbert space by
physical (dynamical) contents then the curvature of this
manifold should have an experimental manifestation.
We shall clarify here some ansatz which has already been used in a latent form in my previous works [2-5]
and will brifly touch upon a possible experimental
test of the evidence of the sectional curvature
in the projective Hilbert space. 

\section{Generalized K\"ahler structure in CP(N)}
Let us introduce {\it generalized K\"ahler structure;
its real part is generalized Fubini-Study metric and
its imaganery part is simplectic structure over 
projective Hilbert space CP(N)}. 

The physical motivation for introduction such  structure
is that arbitrary normalization of wave function is 
acceptable for probabilistic interpretation of quantum
mechanics but this is not sufficient for our purpose--
field foundation of quantum scheme and description of
isolated non-local quantum particles. In particular
in relativistic case there is the problem of normalization because density need not be positive value and this is 
not a probability density but a charge density of field configuration.

Therefore one has some important hints to develop this direction in quantum physics. Namely:

1. The Einstein's formula $E^2=m^2 c^4+c^2 \vec{p}^2$
is source of relativistic wave equations of quantum
theory. But we should remember that this formula was
obtained from the classical point of view in the 
framework of conception of material point. Under 
consistent approach one should obtain this formula 
as a {\it result of classical approximation in quantum
theory}.

2. The hypothesis of field mass together with
Einstein's formula leads to numerical estimation
of spatial extent of quantum particles (classical
radius). But approximate equivalency of classical
radius of quite different kinds of elementary
particles such as electron and proton, for example,
maybe is evidence that there is some unified field 
in the spirit de Broglie--Schr\"odinger--Bohm.

3. Divergences of local quantum theory is artefact 
and consistent theory should not contain these  
divergences. Latests arise under perturbative
approach  in the framework of local linear quantum field theory over such noncompact functional manifold as
ordinary Hilbert space. 
The renormalization process is effectively {\it delocalization of carriers of dynamical variables} and we shall show that this process may be reformuleted 
as a dynamics in a compact manifold of the generalized
coherent states. Here we deal with particular case of
K\"ahler manifold--projective Hilbert space with local coordinates (2.12).

The physical meaning of the metric relationships in the projective Hilbert space has already been discussed
in many works [6-9]. Notwithstanding we should 
return to this question since we propose quite different
physical interpretation of metric, connection and
curvature of this space.

First we will discuss the infinitesimal interval in
separable Hilbert state related to some linear Hermitian
traceless operator $\hat{D} \in AlgSU(N+1)$. This operator creates some interval
$d_{\hat{D}}l^2$ in sense that
\begin{equation}
d_{\hat{D}}l^2=d\Psi_a^* d\Psi^a=<d\Psi|d\Psi>=
<\Psi|\hat{D}^+ \hat{D}|\Psi>d\theta^2,
\end{equation}
where $\theta$ is real-number group parameter. For
instance, if $\hat{D}=\hat{H}$ is Hamiltonian then
one has 
\begin{equation}
d_{\hat{H}}l^2=d\Psi_a^* d\Psi^a=
\hbar^{-2}<\Psi|\hat{H}^+ \hat{H}|\Psi>dt^2.
\end{equation}
We should note now that every nonzero vactor of the
Hilbert space has the isotropy group $H=U(1)\times U(N)$
that leaves this vector intact.  That is transformations which act on state vector effectively lie in the coset
$G/H=SU(N+1)/S[U(1)\times U(N)]$. Furthermore, 
one-parameter transformations in $\theta$ from $G/H$ 
drives state vector along a geodesic in $CP(N)$ [2-5].
Therefore one can transform the scalar product
by the ansatz of ``squeezing'' full state vector to 
the vacuum form [5]
\begin{eqnarray}
<\Psi|\hat{D}^+ \hat{D}|\Psi>=
<\Psi|\hat{G}\hat{G}^{-1}\hat{D}^+\hat{G}\hat{G}^{-1}\hat{D}
\hat{G}\hat{G}^{-1}|\Psi>= \nonumber \\
<0|\hat{D'}(\Psi)^+ \hat{D'}(\Psi)|0>.
\end{eqnarray}
The first ``squeezing'' unitary matrix is
\begin{equation}
\hat{G}_1^+= \left(
\matrix{
1&0&0&.&.&.&0 \cr
0&1&0&.&.&.&0 \cr
.&.&.&.&.&.&. \cr
.&.&.&.&.&.&. \cr
0&.&.&.&1&0&0 \cr
.&.&.&.&0&cos \phi_1&e^{i\alpha_1} sin \phi_1 \cr
0&0&.&.&0&-e^{-i\alpha_1}sin \phi_1&cos \phi_1 
}
\right ). 
\end{equation}
This matrix acts on the general vector
\begin{equation}
|\Psi>= \left(
\matrix{
\Psi^0 \cr
\Psi^1 \cr
. \cr
. \cr
. \cr
\Psi^{N-1}\cr
\Psi^N 
}
\right ) 
\end{equation}
with the result
\begin{equation}
\hat{G}_1^+|\Psi>= \left(
\matrix{
\Psi^0 \cr
\Psi^1 \cr
. \cr
. \cr
. \cr
\Psi^{N-1}cos \phi_1+\Psi^N e^{i\alpha_1} sin \phi_1 \cr
-\Psi^{N-1}e^{-i\alpha_1}sin \phi_1+\Psi^N cos \phi_1 
}
\right ). 
\end{equation}
Now one has solve two ``equations of annihilations'' [5]
$\Re \Psi^{'N}=0$ and $\Im \Psi^{'N}=0$ in order to
eliminate $\Psi^{'N}$ and to find $\alpha'_1$ and $\phi'_1$.
That is one will have a squeezed state vector
\begin{equation}
|\Psi'>= \left(
\matrix{
\Psi^0 \cr
\Psi^1 \cr
. \cr
. \cr
. \cr
\Psi^{N-1}cos \phi'_1+\Psi^N e^{i\alpha'_1} sin \phi'_1 \cr
0
}
\right ). 
\end{equation}
The next step is action of the matrix with the shifted
transformation block
\begin{equation}
\hat{G}_2^+= \left(
\matrix{
1&0&0&.&.&.&0 \cr
0&1&0&.&.&.&0 \cr
.&.&.&.&.&.&. \cr
0&.&.&.&1&0&0 \cr
.&.&.&.&0&cos \phi_2&e^{i\alpha_2} sin \phi_2 \cr
0&0&.&.&0&-e^{-i\alpha_2}sin \phi_2&cos \phi_2 \cr
0&.&.&.&0&0&1 
}
\right ) 
\end{equation}
on the vector (2.7) and evaluation 
$\alpha'_2$ and $\phi'_2$ and so on till the initial 
vector (2.5) will be reduced to the vacuum form 
\begin{equation}
|\Psi_0>= \left(
\matrix{
e^{i\omega} \sum_{a=0}^N |\Psi^a|^2 \cr
0 \cr
. \cr
. \cr
. \cr
0 
}
\right ). 
\end{equation}
That is $|\Psi_0>=\hat{G}^{-1}|\Psi>$, where 
$\hat{G}=\hat{G}_1 \hat{G}_2...\hat{G}_N$.
It is clear that this process of ``squeezing'' is 
equivalent
to the reduction of quadric (2.1) to main axes as
\begin{equation}
d_{\hat{D}}l^2=<\Psi_0|\hat{D'}(\Psi)^+ \hat{D'}(\Psi)|\Psi_0>=
R^2(|D'_{00}(\Psi)|^2+\sum_{i=1}^N |D'_{0i}(\Psi)|^2)d\theta^2.
\end{equation}
This contains two part: the first term is interval along
the subalgebra of the isotropy group of the vector
$|\Psi_0>$ and the second one is interval in the 
tangent space to the coset itself.
We saw that diagonalization of the quadric (2.1)
of the infinitezimal interval in the state space which related to some dynamical variable $\hat{D}$, connected
with the coset structure $G/H=SU(N+1)/S[U(1) \times U(N)]$.
That is here the full unitary symmetry $SU(N+1)$ has 
become spontaneously broken down to the isotropy group 
$U(1) \times U(N)$. 
From the topological point of view $G/H$-structure is equivalent to the structure of the projective Hilbert 
space $CP(N)$ [10]. This paves the way to the invariant
study of the spontaneously broken unitary symmetry.
Elements of tangent space to the coset will be functions 
of state vector  during ansatz of ``squeezing'' . In that
sense local (in a functional space) dynamical variables 
arise. But invariant properties of the interval should be
independent of a choice of the dynamical variable 
$\hat{D}$. There is a direct method of introduction 
of the local dynamical variables in the projective Hilbert 
space in local coordinates [2-5]. It corresponds the
well known ``active'' point of view on transformations
of state vectors. Now we looking for invariant properties
of  the infinitezimal interval $\delta L^2$ in the original Hilbert space as local dynamical variables have already 
been discussed [2-5]. 

We start with ordinary decomposition of a state vector
of quantum system in some orthogonal basis
$|\Psi>=\sum_{a=0}^N \Psi^a|a>$ where 
$\sum_{a=0}^N |\Psi^a|^2=R^2$,  $(0\leq a \leq N)$
or 
$|\Psi>=\sum_{a=0}^{\infty}\Psi^a|a>$ where 
$\sum_{a=0}^{\infty}|\Psi^a|^2=R^2$,  $(0\leq a < \infty)$.
The generalized stereographic projection from the center
of the sphere $\sum_{a=0}^N |\Psi^a|^2=R^2$ onto the 
complex hyperplane $\Pi$ give us relationships between
coordinates of a point of the sphere in the original 
Hilbert space $\Psi^0,...,\Psi^a,...,\Psi^N$ and 
coordinates
$\Pi^1,...,\Pi^i,...,\Pi^N$
of its projection onto the hyperplane
\begin{equation}
\Psi^0=\lambda(R,\Pi)R,\quad \Psi^1=\lambda(R,\Pi)\Pi^1,
\Psi^2=\lambda(R,\Pi)\Pi^2,...,\Psi^N=\lambda(R,\Pi) \Pi^N,...
\end{equation}
Then one has ($1 \leq i \leq N$ or $1 \leq i < \infty$)
\begin{equation}
\Pi^1=R\frac{\Psi^1}{\Psi^0},\quad \Pi^2=R\frac{\Psi^2}{\Psi^0},...,
\Pi^i=R\frac{\Psi^i}{\Psi^0},...,
\Pi^N=R\frac{\Psi^N}{\Psi^0},...
\end{equation}
and
$\lambda^2 (\sum_{i=1}^N |\Pi^i|^2+R^2)=R^2$
or 
$\lambda^2 (\sum_{i=1}^{\infty}|\Pi^i|^2+R^2)=R^2$.
The ``squeezing factor'' $\lambda(R,\Pi)$ one can 
express from these equations
\begin{equation}
\lambda(R,\Pi)=\frac{R}{\sqrt{\sum_{s=1}^N |\Pi^s|^2+R^2}}
\end{equation}
or, for $N \to \infty$
\begin{equation}
\lambda(R,\Pi)=\frac{R}{\sqrt{\sum_{s=1}^{\infty} |\Pi^s|^2+R^2}}.
\end{equation}
Hereafter we will use only finite value of indexes 
$0 \leq a,b,...,d \leq N$ and $1 \leq i,k,...,s \leq N $
remembering that in typical cases the limit $N \to \infty $
may be achieved.
Now we can express homogeneous coordinates $\Psi$ in local
coordinates $\Pi$:
\begin{equation}
\Psi^0=\frac{R^2}{\sqrt{\sum_{s=1}^N |\Pi^s|^2+R^2}},...,
\quad 
\Psi^i=\Pi^i \frac{R}{\sqrt{\sum_{s=1}^N |\Pi^s|^2+R^2}},...
\end{equation}
It is easy to evaluate ($a=0$)
\begin{equation}
\frac{\partial \Psi^0}{\partial \Pi^i}=-\frac{1}{2}
\frac{R^2 \Pi^{*i}}{\left(\sqrt{\sum_{s=1}^N |\Pi^s|^2+R^2}\right)^3},
\frac{\partial \Psi^0}{\partial \Pi^{*k}}=-\frac{1}{2}
\frac{R^2 \Pi^{k}}{\left(\sqrt{\sum_{s=1}^N |\Pi^s|^2+R^2}\right)^3}
\end{equation}
and for other components ($a \geq 1$) one has
\begin{eqnarray}
\frac{\partial \Psi^a}{\partial \Pi^i}     & = &
R\left(\frac{\delta^a_i}{\sqrt{\sum_{s=1}^N |\Pi^s|^2+R^2}}-
\frac{1}{2} \frac{\Pi^a \Pi^{*i}}
{\left(\sqrt{\sum_{s=1}^N |\Pi^s|^2+R^2}\right)^3}\right), \nonumber \\
\frac{\partial \Psi^{*a}}{\partial \Pi^{*k}} & = &
R\left(\frac{\delta^a_k}{\sqrt{\sum_{s=1}^N |\Pi^s|^2+R^2}}-
\frac{1}{2} \frac{\Pi^{*a} \Pi^{k}}
{\left(\sqrt{\sum_{s=1}^N |\Pi^s|^2+R^2}\right)^3}\right). 
\end{eqnarray}
Therefore one can express infinitezimal invariant interval
in the original Hilbert space as followes
\begin{equation}
\delta L^2= \delta_{ab}\delta \Psi^a \delta \Psi^{*b}= G_{ik*}\delta \Pi^i \delta \Pi^{*k}=\sum_a \frac{\partial \Psi^a}{\partial \Pi^i}
\frac{\partial \Psi^{*a}}{\partial \Pi^{*k}} 
\delta \Pi^i \delta \Pi^{*k}.
\end{equation}
That is the generalized metric tensor of the original flat
Hilbert space in the local coordinates $\Pi$ is 
\begin{eqnarray}
G_{ik*}=\sum_{a=0}^N \frac{\partial \Psi^a}{\partial \Pi^i}
\frac{\partial \Psi^{*a}}{\partial \Pi^{*k}} & = &\nonumber \\
R^2 \left[\frac{(\sum_{s=1}^N |\Pi^s|^2+R^2)\delta_{ik}-
\Pi^{*i}\Pi^k}{(\sum_{s=1}^N |\Pi^s|^2+R^2)^2}+ \frac{1}{4}\frac{(\sum_s^N |\Pi^s|^2 +R^2)\Pi^{*i}\Pi^k}
{(\sum_{s=1}^N |\Pi^s|^2+R^2)^3}\right] & = & \nonumber \\
R^2 \frac{(\sum_{s=1}^N |\Pi^s|^2+R^2)\delta_{ik}-
\frac{3}{4}\Pi^{*i}\Pi^k}{(\sum_{s=1}^N |\Pi^s|^2+R^2)^2}.
\end{eqnarray}
This metric tensor contains two parts: the first one
which arises from the geometry of the projective Hilbert  
space $CP(N)$;
if all $\Pi^i$ are small in comparison with $R$ then the
term
\begin{equation}
\frac{R^2}{4}\frac{\Pi^{*i} \Pi^k \sum_s^N |\Pi^s|^2}
{(\sum_{s=1}^N |\Pi^s|^2+R^2)^3}
\end{equation}
is negligible. In this case one has the commonly known Fubini-
Study metric tensor for $R=1$ [10]. In the general case  this term should be taken into account. The second part 
arises from the derivatives of the component $\Psi^0$ 
which is orthogonal to the hyperplane $\Pi$.
\section{Discussion}
It is clear now why so difficult to see the evidence 
of the
curvature of the projective Hilbert space in physical
experiment. For small $\Pi^i$ the full 
invariant interval
\begin{eqnarray}
\delta L^2=R^2\frac{(\sum_{s=1}^N |\Pi^s|^2+R^2)\delta_{ik}-
\frac{3}{4}\Pi^{*i}\Pi^k}{(\sum_{s=1}^N |\Pi^s|^2+R^2)^2}
\end{eqnarray}
in original Hilbert space is  numerically very close to 
the interval
\begin{eqnarray}
dl^2=R^2\frac{(\sum_{s=1}^N |\Pi^s|^2+R^2)\delta_{ik}-
\Pi^{*i}\Pi^k}{(\sum_{s=1}^N |\Pi^s|^2+R^2)^2}
\end{eqnarray}
in the projective Hilbert space $CP(N)$. Notwithstanding
we think that
it is possible to find a trace of the non-zero sectional curvature of $CP(N)$
in optic experiments for the measuring of Aharonov-Anandan
phase for light rays with different parameters. 
This should be truly ``geometric phase'' effect in distinction from all previous ``topological phase'' 
effects as it depends not only on a topology of the 
way in the projective Hilbert space but on the metric properties of the latest as well. 
It will be discussed together with the physical meaning 
of the radius $R$ of the sectional curvature $1/R^2$ 
elsewhere.

\bigskip
                          
ACKNOWLEDGEMENTS

\bigskip
I sincerely thank Lawrence P.Horwitz for useful discussions.

\end{document}